\documentclass[conference]{IEEEtran}

\IEEEoverridecommandlockouts
\usepackage{cite}
\usepackage{amsmath,amssymb,amsfonts}
\usepackage{algorithmic}
\usepackage{graphicx}
\usepackage{textcomp}
\usepackage{xcolor}
\usepackage{subcaption}
\def\BibTeX{{\rm B\kern-.05em{\sc i\kern-.025em b}\kern-.08em
    T\kern-.1667em\lower.7ex\hbox{E}\kern-.125emX}}
    
\usepackage{fancyhdr}
{
\fancyhead[C]{Published as a conference paper at IEEE TPS 2019}

\begin{document}
\setlength{\abovedisplayskip}{0pt}
\setlength{\belowdisplayskip}{0pt}
\title{Twitter Bot Detection Using Bidirectional Long Short-term Memory Neural Networks and Word Embeddings\\}

\author{\IEEEauthorblockN{Feng Wei and Uyen Trang Nguyen}
\IEEEauthorblockA{{Department of Electrical Engineering and Computer Science} \\
{York University}\\
4700 Keele Street, Toronto, Canada M3J 1P3 \\
\{fwei, utn\}@eecs.yorku.ca}
}

\maketitle

\begin{abstract}
Twitter is a web application playing dual roles of online social networking and micro-blogging. The popularity and open structure of Twitter have attracted a large number of automated programs, known as bots. Legitimate bots generate a large amount of benign contextual content, i.e., tweets delivering news and updating feeds, while malicious bots spread spam or malicious contents. To assist human users in identifying who they are interacting with, this paper focuses on the classification of human and spambot accounts on Twitter, by employing recurrent neural networks, specifically bidirectional Long Short-term Memory (BiLSTM), to efficiently capture features across tweets. To the best of our knowledge, our work is the first that develops a recurrent neural model with word embeddings to distinguish Twitter bots from human accounts, that requires no prior knowledge or assumption about users' profiles, friendship networks, or historical behavior on the target account. Moreover, our model does not require any handcrafted features. The preliminary simulation results are very encouraging. Experiments on the cresci-2017 dataset show that our approach can achieve competitive performance compared with existing state-of-the-art bot detection systems.
\end{abstract}


\section{INTRODUCTION}
Twitter is a popular online social networking and micro-blogging
tool. Remarkable simplicity is its distinctive feature: its community interacts via publishing text-based posts, known as tweets. Twitter has its own special memes, i.e., hashtag (\#), mention (@), shortened URL (http://t.co) and retweet (RT). Hashtags, namely words or phrases prefixed with a \# symbol, can group tweets by topics. For example, \#\emph{usopen2019} and \#\emph{SheTheNorth} are two trending hashtags on Twitter in September 2019. The symbol @ followed by a user name in a tweet enables the direct delivery of the tweet to that user. Links shared on Twitter, including links shared in private direct messages, will automatically be processed and shortened to an http://t.co link. A retweet is a re-posting of a tweet. Sometimes people type ``RT" at the beginning of a Tweet to indicate that they are re-posting someone else's content.

The growing user population and open nature of Twitter have made itself an ideal target of exploitation from automated programs, known as bots. Automation is a double-edged sword to Twitter. On the one hand, legitimate bots generate a large volume of benign tweets, like news and blog updates. On the other hand, malicious bots have been widely exploited to spread spam or malicious contents. The definition of spam in this paper is to spread malicious, phishing, or unsolicited content in tweets. These bots randomly follow human users, expecting many users to follow them back.

The spambot problem in social networks has already received attention from researchers. As an example for spam detection, a branch of research mined the textual content of tweets \cite{gao2012towards}; others studied the redirection of embedded URLs in tweets \cite{lee2013warningbird}, or classified the URLs landing pages \cite{thomas2011design}. Gao et al. \cite{gao2014spam} move beyond the difficulty of labeling those tweets without URLs as spam tweets, by proposing a composite tool able to match incoming tweets with underlying templates commonly used.  Cresci et al. \cite{cresci2016dna} introduced a bio-inspired technique to model online user behaviors. 

Most existing work identify spambots through a multi-feature approach, including features on the profile, user behavior, friendship networks, and the timeline of an account. Examples of such analyses include \cite{yang2013empirical,lee2011seven,davis2016botornot,alsaleh2014tsd,varol2017online,ahmed2013generic,miller2014twitter}. In addition, Yang et al. \cite{yang2013empirical} designed a series of new criteria, and demonstrated their effectiveness in detecting spambots that evade previous detection techniques.

In this paper, we propose a recurrent neural network (RNN) model, specifically BiLSTM, with word embeddings to distinguish Twitter bots from human accounts.  To the best of our knowledge, our work is the first that develops a RNN model with word embeddings to detect Twitter bots that requires no prior knowledge or assumption about users' profiles, friendship networks, or historical behavior on the target account.

We summarize our contributions as follows.
\begin{itemize}
\item We propose a RNN model to distinguish Twitter bots from human accounts, instead of using traditional methods (e.g., Random Forest, Bayes Net or Support Vector Machine). BiLSTM connects two hidden layers of opposite directions to the same output. With this form of generative deep learning, the output layer can get information from past (backwards) and future (forward) states simultaneously. This method can efficiently capture contextual features across tweets and achieve competitive classification accuracy compared to existing methods.
\item We use word embeddings to encode tweets, instead of using traditional feature engineering or natural language processing (NLP) tools that are much more complex. This advantage allows for faster and easier implementation and deployment of the bot detection scheme.
\item We conducted experiments on real-world data sets. Experiments on the cresci-2017 dataset show that our approach can achieve competitive performance compared with existing work~\cite{cresci2016dna,yang2013empirical,davis2016botornot,ahmed2013generic,miller2014twitter}, although our RNN model uses only the contextual content of tweets as the input to the model.
\end{itemize}

The remainder of the paper is organized as follows. We discuss related work in Section II. The proposed approach is described in Section III. In Section IV, we present experimental results, comparing the performance of our proposed model with that of existing state-of-the-art systems. Section V concludes the paper and outlines our future work.

\section{RELATED WORK}
We discuss related work on existing techniques of Twitter bot detection and recurrent neural networks.

\subsection{Twitter Bot Detection}
Traditional bot detection systems typically rely on the application of well-known machine learning algorithms on the accounts under investigation, such as \cite{yardi2010detecting,benevenuto2010detecting,ghosh2012understanding,lee2010uncovering,stringhini2010detecting,
viswanath2011analysis,thomas2011design,gao2012towards,cao2012aiding,xie2012innocent,yang2013empirical,wang2012social,lee2013warningbird,yang2014uncovering,liu2014sdhm,
paradise2014anti,cresci2014criticism,ferrara2016rise,botornot}. However, since 2013, a number of research teams independently started to formalize new approaches for detecting the coordinated and synchronized behavior that characterizes groups of automated malicious accounts~\cite{beutel2013copycatch,giatsoglou2015nd,jiang2016catching,cao2014uncovering,viswanath2015strength,cresci2016dna}. Despite being based on different key concepts, these studies investigate groups of accounts as a whole, marking a difference with the previous literature. However, our approaches for the spam problem focus on the detection of tweets containing spam instead of detecting spam accounts. The detection of spam tweets itself can be useful for filtering spam on real time search~\cite{benevenuto2010detecting}, whereas the detection of spammers is related with the detection of existent spam accounts. In fact, a way to detect spammers would be filtering users who have written many spam tweets. In addition, when a spam account is detected, Twitter suspends it or even blocks his IP address temporally, so spammers only need to create a different account to continue sending spam messages or wait a while for his IP address is unlocked.

Existing methods of Twitter bot detection can be divided into two main approaches:  supervised machine learning  and unsupervised clustering. Both of them require complex handcrafted features. \\
\textbf{Supervised machine learning strategy.} Lee et al. \cite{lee2011seven} adopt 30 classification algorithms and tested their performance. Tree-based supervised classifiers showed the highest accuracy results. In particular, Random Forest produced the highest accuracy. In order to improve the Random Forest classifier, standard boosting and bagging techniques have been  applied additionally. The authors trained the content polluter classifier based on different feature group combinations. The system in \cite{yang2013empirical} provides a supervised machine learning classifier that infers whether a Twitter account is a human account or a spambot by relying on relationships among accounts, tweeting timing and level of automation. In addition, they design 10 new behavior detection features. According to their evaluation, the detection rate using their new feature set is significantly higher than that of existing work. Alsaleh et al. \cite{alsaleh2014tsd} presented a system that utilizes supervised machine learning techniques to dynamically detect Twitter bot accounts. The classification results show satisfying detection rate for this particular application. Although they adopt multilayer neural network, which is a simple feedforward neural network (FFNN), it still require complicated handcrafted features. Davis et al. \cite{davis2016botornot} group features into six main classes: network, user, friend, temporal, content and sentiment, and employ Random Forest, an ensemble supervised learning method to achieve a high accuracy score. Varol et al. \cite{varol2017online} proposed a supervised machine learning system that extracts more than a thousand features in six different classes: users, friends meta-data, tweet content, sentiment, network patterns and activity time series.\\
\textbf{Unsupervised clustering approach.} The approach in \cite{miller2014twitter} considers vectors made of 126 features, extracted from both accounts and tweets, as input to modified versions of the DenStream \cite{cao2006density} and StreamKM++ \cite{ackermann2012streamkm++} clustering algorithms, to cluster feature vectors of a set of unlabeled accounts. The methodology in \cite{ahmed2013generic} exploits a set of 14 generic statistical behavior features related to URLs, hashtags, mentions and retweets. Feature vectors generated in this way are then compared with one another via an Euclidean distance measure. Chavoshi et al. \cite{chavoshi2016debot} developed an unsupervised method, named DeBot, which calculates cross-user activity correlations to detect bot accounts in Twitter. Debot detects thousands of bots per day with a 94\% precision and generates reports online everyday. Cresci et al. \cite{cresci2016dna} proposed an unsupervised method to detect spambots, by comparing their behavior with the aim of finding similarities between automated accounts. They introduced a bio-inspired technique to model online user behaviors by so-called ``digital DNA" sequences. Extracting digital DNA for an account means associating that account to a string that encodes its behavioral information. Although it achieves good detection performances, numerous handcrafted behavioral features are still required.

A recent research direction is to test the limits of current bot detection frameworks in an adversarial setting. The idea is to propose methodologies to engineer systems that can go undetected. Cresci et al.~\cite{cresci2019capability} proposed the use of evolutionary algorithms to improve social bot skills. Grimme et al.~\cite{grimme2018changing} employed a hybrid approach involving automatic and manual actions to create bots that would be classified as human by a supervised bot detection system. Despite the good intention of pointing to weaknesses in existing systems, this research might also inspire bot creators and give them a competitive advantage.

\subsection{Recurrent Neural Networks}
In the past few years, deep neural networks have achieved huge successes in many data modelling and prediction tasks, ranging from speech recognition, computer vision to natural language processing (NLP). In this paper, we  apply powerful deep learning methods to social network data modelling to distinguish Twitter bots from human accounts.

Deep learning approaches are able to automatically capture, to some extent, the syntactic and semantic features from contextual content without handcrafted feature engineering, which is labor intensive and time consuming. They attract much research interest in recent years, and achieve state-of-the-art performances in many fields of NLP.

Socher et al. \cite{socher2013recursive} first propose a family of recurrent neural networks (RNN) to learn the compositional semantic of variable-length phrases and sentences. Irsoy et al. \cite{irsoy2014deep} present a deep RNN constructed by stacking multiple recurrent layers for compositionality in language. Long short-term memory (LSTM) \cite{hochreiter1997long}, which is a kind of RNN architecture, is explicitly designed to solve the long-term dependency problem through purpose-built memory cells. BiLSTM \cite{graves2013speech} incorporates a forward LSTM layer and a backward LSTM layer, in order to learn information from preceding as well as following tokens.

In this paper, we define the problem of Twitter bot detection as a text classification problem: we use only the contextual content of tweets as the input to our RNN model.

\section{OUR PROPOSED APPROACH}
In this section, we discuss our proposed method of bidirectional LSTM (BiLSTM) with word embeddings.

\subsection{Word Embeddings}
Human vocabulary comes in free text. In order to make a machine learning model understand and process the natural language, we need to transform the free-text words into numeric values. One of the simplest transformation approaches is to do a one-hot encoding, in which each distinct word stands for one dimension of the resulting vector and a binary value indicates whether the word presents (one) or not (zero).

Word embedding is a dense representation of words in the form of numeric vectors. It can be learned using a variety of language models. The most exciting point from word embedding is that, similar words are located together in the vector space, and arithmetic operations on word vectors can pose semantic or syntactic relationships. For example, vector ``cat" - vector ``kitten" is similar to vector ``dog" - vector ``puppy". However, traditional machine learning approach (e.g., Latent Dirichlet Allocation) cannot maintain such a linear relationship in the vector space.

The Global Vector (GloVe) model, proposed by Pennington et al. \cite{pennington2014glove} aims to combine the count-based matrix factorization and the context-based skip-gram model together. In other words, the motivation of GloVe is to force the model to learn such linear relationship based on the co-occurreence matrix explicitly. Essentially, GloVe is a log-bilinear model with a weighted least-squares objective. Obviously, it is a hybrid method that uses machine learning based on the statistic matrix.

Word embeddings, also known as distributed word representation, is an important research topic in NLP. In recent years, it has been widely used in various NLP task, including information retrievals~\cite{xu2016yorknrm,xu2017fofe,wei2019dual}, text classification \cite{wang2019investigating}, machine translation \cite{clinchant2019use} and machine comprehension \cite{xu2019ranking}. The success of word embedding \cite{mikolov2013distributed,pennington2014glove} encourages researchers to focus on machine-learned representation instead of heavy feature engineering in NLP. By using word embeddings as the typical feature representation for words, neural networks become more competitive compared to traditional approaches in NLP.


An important advantage of word embeddings compared to conventional NLP techniques of representation (e.g., bag-of-words~\cite{harris1954distributional}, part-of-speech tagging~\cite{jurafsky2000speech}) is that it achieves a significant dimensional reduction of the feature set needed to represent tweets, resulting in a reduction in training and inference time of the algorithms.

In this work, we adopt pre-trained GloVe word vectors on Twitter. It is based on 2 billion tweets, including 27 billion tokens, with the vocabulary size of 1.2 million. We define our vocabulary as the intersection between the words in all training samples and those in the pre-trained 200-dimensional GloVe. Given a word $w$, if it is in the vocabulary, we set its word-level embedding $\alpha_w$ to its GloVe word vector, which is fixed during the training; otherwise we have $\alpha_w = \alpha_o \in \mathbb{R}^{200}$, where $\alpha_o$ is a trainable parameter serving as the shared word vector of all out-of-vocabulary (OOV) words. Each tweet is sent to the Stanford CoreNLP \cite{manning2014stanford} toolkit for sentence splitting and tokenization. All words containing Twitter special memes, i.e., hashtag (\#), mention (@) and shortened URL (http://t.co), are mapped to several pre-defined tokens individually, i.e., $\langle HASHTAG\rangle$, $\langle USER\rangle$ and $\langle URL\rangle$, using regular expression matches.

\subsection{BiLSTM Neural Networks}
First, we briefly describe RNN, LSTM and BiLSTM individually. Then, our proposed model is described.

\begin{figure}[htbp]
\center{\includegraphics[height=3cm]{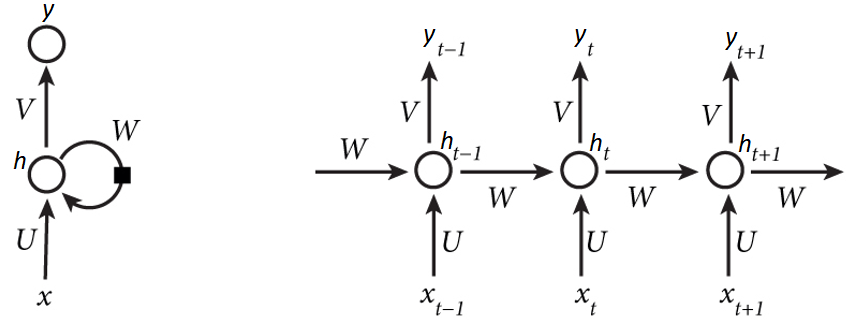}}
\caption{\textbf{Left}: Recurrent neural network; \textbf{Right}: Recurrent neural network unfold.}
\label{fig:1}
\end{figure}

\textbf{RNN} \cite{elman1990finding} is a class of artificial neural sequence model, as shown in Fig.~\ref{fig:1}, where connections between units form a directed cycle. It takes arbitrary embedding sequences $x = (x_1, ...  , x_T)$ as input, uses its internal memory network to exhibit dynamic temporal behavior. It consists of a hidden unit $h$ and an optional output $y$. $T$ is the last time step and is also the length of input sentence in this text sequence learning task. At each time step $t$, the hidden state $h_t$ of the RNN is computed based on the previous hidden state $h_{t-1}$ and the input at the current step $x_t$:
\begin{equation}
h_t = g(Ux_t + Wh_{t-1}) \label{eq:1}
\end{equation}
where $U$ and $W$ are weight matrices of the network; $g(\cdot)$ is a non-linear activation function, such as an element-wise logistic sigmoid function. The output at time step $t$ is computed as $y_t = softmax(Vh_t)$, where $V$ is another weight parameter of the network; $softmax$ is an activation function often implemented at the final layer of a network.

\begin{figure}[htbp]
\center{\includegraphics[height=5cm]{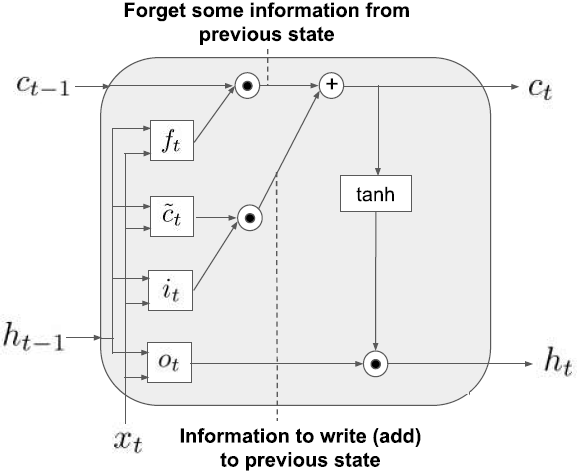}}
\caption{Long Short-term Memory Cell.}
\label{fig:2}
\end{figure}

\textbf{LSTM} \cite{hochreiter1997long} is a variant of RNN designed to deal with vanishing gradients problem. However, Hochreiter and Schmidhuber~\cite{hochreiter1997long} found that the proposed architecture, which uses purpose-built memory cells to store information, is better at finding and exploiting long range context. Fig.~\ref{fig:2} illustrates a single LSTM memory cell.  For the version of LSTM used in \cite{gers2002learning}, $g(\cdot)$ is implemented by the following composite function:
\begin{gather}
i_t = \sigma (U_ix_t + W_ih_{t-1} + V_ic_{t-1}) \label{eq:2} \\
f_t = \sigma (U_fx_t + W_fh_{t-1} + V_fc_{t-1}) \label{eq:3} \\
c_t = f_{t}c_{t-1} + i_t\ tanh(U_cx_t + W_ch_{t-1}) \label{eq:4} \\
o_t = \sigma (U_ox_t + W_oh_{t-1} + V_oc_t) \label{eq:5} \\
h_t = o_t\ tanh(c_t) \label{eq:6}
\end{gather}
where $\sigma$ is the logistic sigmoid function, and $i$, $f$, $o$ and $c$
are the $input~gate$, $forget~gate$, $output~gate$ and $cell$ activation vectors, respectively, all of which are of the same size as the hidden vector $h$; $U$, $W$ and $V$ are weight matrices of the network. The weight matrices from the cell to gate vectors (e.g., $W_i$) are diagonal, so element $m$ in each gate vector only receives input from element $m$ of the cell vector.

\begin{figure}[htbp]
\center{\includegraphics[height=4cm]{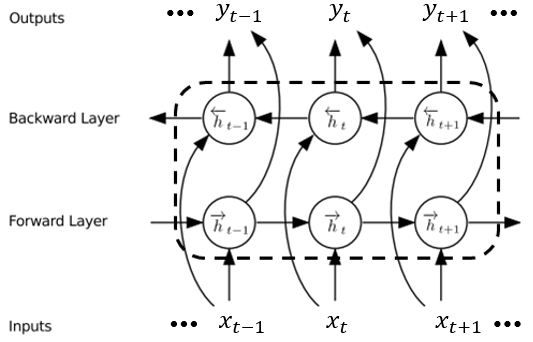}}
\caption{Bidirectional LSTM.}
\label{fig:3}
\end{figure}

\textbf{BiLSTM} uses two LSTMs to learn each token of the sequence based on both the past and the future context of the token. As shown in Fig.~\ref{fig:3}, one LSTM processes the sequence from left to right; the other one from right to left. At each time step $t$, a hidden forward layer with hidden unit function $\overrightarrow{h}$ is computed based on the previous hidden state $\overrightarrow{h}_{t-1}$ and the input at the current step $x_t$. Additionally, a hidden backward layer with hidden unit function $\overleftarrow{h}$ is computed based on the future hidden state $\overrightarrow{h}_{t+1}$ and the input at the current step $x_t$. The forward and backward context representations, generated by $\overrightarrow{h}_t$ and $\overleftarrow{h}_t$ respectively, are concatenated into a long vector. The combined outputs are the predictions of target sequences.

\begin{figure}[htbp]
\center{\includegraphics[height=8cm]{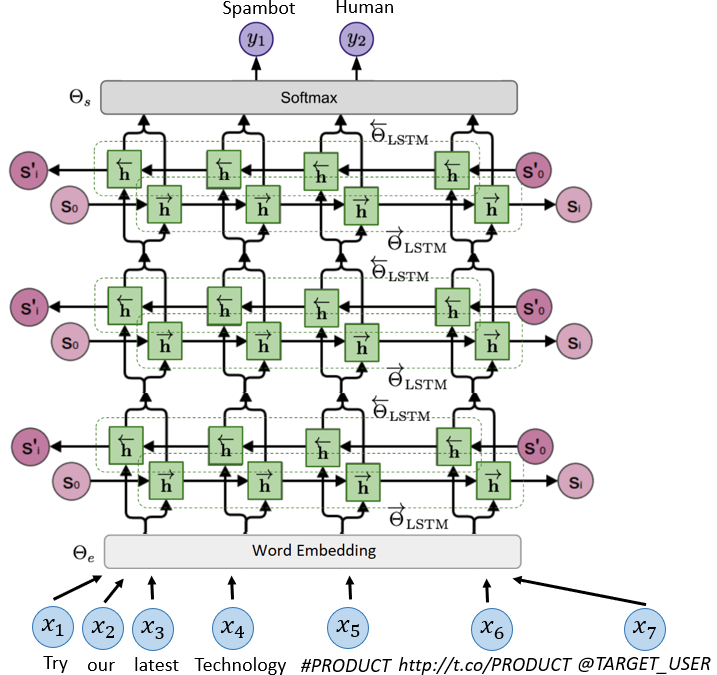}}
\caption{An illustration of our BiLSTM neural model with word embeddings for distinguishing Twitter bots from human accounts.}
\label{fig:4}
\end{figure}

\textbf{Our Proposed Model}. As shown in Fig.~\ref{fig:4}, we make use of a fully connected softmax layer to output posterior probabilities over labels from two classes, standing for Twitter bots or human. The input is a sequence of $n$ tokens, ($x_1, ..., x_n$). The predictions in both directions are modeled by three-layer BiLSTMs with hidden states $\overrightarrow{h}_{i,\ell}$ and $\overleftarrow{h}_{i,\ell}$ for input token $x_i$ at the layer level $\ell=\emph{1, ..., L}$. The final layer's hidden state $h_{i,L}=[\overrightarrow{h}_{i,L};\overleftarrow{h}_{i,L}]$ is used to output the probabilities over binary labels after softmax normalization. They share the word embedding layer and the softmax layer, parameterized by $\Theta_e$ and $\Theta_s$ respectively. The model is trained to minimize the negative log likelihood in both directions:
\begin{multline}
\mathcal{L}=-\sum_{i=1}^{n}(logp(y|x_1, ..., x_n;\Theta_e,\overrightarrow{\Theta}_{LSTM},\Theta_s) + \\
logp(y|x_1, ..., x_n;\Theta_e,\overleftarrow{\Theta}_{LSTM},\Theta_s))
\label{eq:7}
\end{multline}

\section{EXPERIMENTAL RESULTS}
In this section, we present our experimental setup and results, comparing the performance of our neural model with that of existing work~\cite{cresci2016dna,yang2013empirical,davis2016botornot,ahmed2013generic,miller2014twitter}.

\subsection{Existing Systems for Comparison}
\begin{figure}[htbp]
\center{\includegraphics[height=4cm]{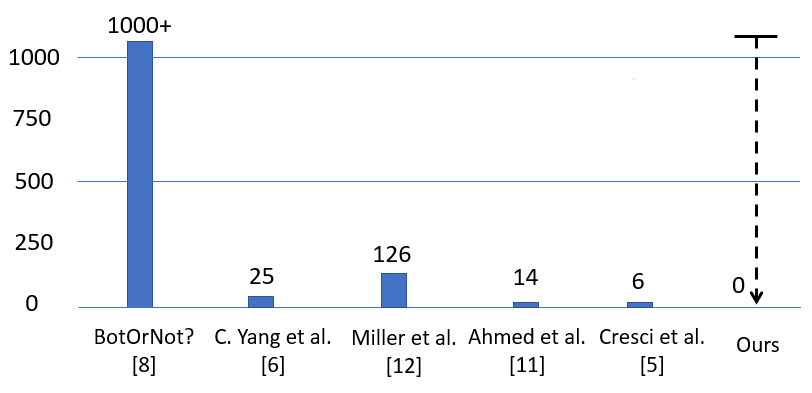}}
\caption{The number of handcrafted feature comparison among various spambot detection techniques and algorithms reported on the cresci-2017 dataset.}
\label{fig:feature}
\end{figure}

Davis et al. \cite{davis2016botornot} generate more than 1,000 features and group them into six main classes: network, user, friend, temporal, content and sentiment. Yang et al. \cite{yang2013empirical} use 25 features and group them into six categories: profile-based features, content-based features, graph-based features, neighbor-based features, timing-based features and automation-based features. Miller et al. \cite{miller2014twitter} consider vectors made of 126 features, extracted from both accounts and tweets. Ahmed et al. \cite{ahmed2013generic} exploit a set of 14 generic statistical behavior features related to URLs, hashtags, mentions and retweets. Cresci et al. \cite{cresci2016dna} design two groups of user behaviour features: tweet type DNA and tweet content DNA.  It is worth noting that most   state-of-the-art algorithms/systems for spambot detection require a large number of data-demanding features. As shown in Fig.~\ref{fig:feature}, the existing work requires feature engineering based on six~\cite{cresci2016dna} to more than 1000 features~\cite{davis2016botornot}.   Feature engineering is very expensive in terms of data collection, pre-processing and computation power.   Our proposed RNN model does not rely on any feature engineering and uses  only the contextual content of the tweets.

\subsection{Dataset}
We evaluate our proposed model using the public annotated dataset cresci-2017~\cite{mibdatasets}, consisting of 3,474 human accounts along with 8.4 million tweets and 1,455 bots along with 3 million tweets. We prepared  two test sets  following \cite{cresci2017paradigm}. Test set \#1 and test set \#2 refer to groups where human accounts are mixed with accounts from dataset social-bot-1 and dataset social-bot-3, respectively. Social-bot-1 is about retweeters of an Italian political candidate, while social-bot-3 is about spammers of products on sale at Amazon.com. Test set \#1 is composed of 1,982 accounts and 4,061,598 tweets, while test set \#2 is composed of 928 accounts and 2,628,181 tweets. The statistics of datasets are shown in Table \ref{tb:dataset} in detail.

\begin{table*}[t!]
\caption{Statistics about the datasets used for this work.}
\begin{center}
  \begin{tabular}{c c c c}
  \hline      
   & & \multicolumn{2}{c}{\textbf{statistics}} \\
  \cline{3-4}      
\textbf{dataset}  &  \textbf{description}  & accounts &  tweets  \\
\hline      
human accounts            & verified accounts that are human-operated            & 3,474 &  8,377,522	          \\
social-bot-1              & retweeters of an Italian political candidate         & 991   &  1,610,176	          \\
social-bot-3              & spammers of products on sale at Amazon.com           & 464   &  1,418,626	          \\
\hline      
test set \#1              & mixed set of 50\% human accounts + 50\% social-bot-1 & 1,982 &  4,061,598             \\
test set \#2              & mixed set of 50\% human accounts + 50\% social-bot-3 & 928   &  2,628,181             \\
\hline      
\end{tabular}
\label{tb:dataset}
\end{center}
\end{table*}

\subsection{Neural Network Model Setup}
Based on the design of the experiments, we tested several sets of parameters to select one that gives the experiments the best performance.  These parameters are as follows:

\begin{itemize}
    \item
\emph{Learning rate}: the model is trained using the Stochastic Gradient Descent algorithm, while the learning rate is set to  0.01.
\item \emph{Network structure}: three stacked BiLSTM layers with 200 recurrent units and one fully connected softmax layer.
\item   \emph{Dropout} \cite{hinton2012improving} is adopted during training, initially set to 0.5, slowly decreasing during training until it reaches 0.1 at the end.
\item \emph{Number of epochs}: 30.
\item \emph{Momentum}: 0.9.
\item \emph{Mini-batch}: 64.
\end{itemize}

\subsection{Evaluation Metrics}
To evaluate the effectiveness of our proposed RNN approach, we use four standard indicators:
\begin{itemize}
\item \emph{True Positives (TP)}: the number of spambots correctly recognized;
\item \emph{True Negatives (TN)}: the number of human accounts correctly recognized;
\item \emph{False Positives (FP)}: the number of human accounts erroneously recognized as spambots;
\item \emph{False Negatives (FN)}: the number of spambots erroneously recognized as human accounts.
\end{itemize}

For each test set, we use the following standard evaluation metrics to compare the performance of the classifiers:
\begin{itemize}
\item \emph{Precision}, the ratio of predicted positive cases, i.e., Twitter bots, that are indeed real positives: $\frac{TP}{TP+FP}$;
\item \emph{Recall} (also known as Sensitivity), the ratio of real positive cases that are indeed predicted as positives: $\frac{TP}{TP+FN}$;
\item \emph{Specificity}, the ratio of real negative cases, i.e., human accounts, that are correctly identified as negative: $\frac{TN}{TN+FP}$;
\item \emph{Accuracy}, the ratio of correctly classified users (both positives and negatives) among all the users: $\frac{TP+TN}{TP+TN+FP+FN}$;
\item \emph{F-measure}, the harmonic mean of Precision and Recall: $2\cdot\frac{Precision\cdot Recall}{Precision+Recall}$;
\item \emph{Matthews Correlation Coefficient (MCC)} \cite{baldi2000assessing}, the estimator of the correlation between the predicted class and the real class of the users: $\frac{TP\cdot TN-FP\cdot FN}{\sqrt{(TP+FN)\cdot(TP+FP)\cdot(TN+FP)\cdot(TN+FN)}}$\\
\end{itemize}

Each of the above metrics captures a different aspect of the prediction performance. \emph{Accuracy} measures how many users are correctly classified in both of the classes, but it does
not express whether the positive class is better recognized than the other one. Furthermore, there are situations where some predictive models perform better than others, even having a lower accuracy. A high \emph{Precision} indicates that many of the users identified as spambots are indeed real spambots, but it does not give any information about the number of spambots that have not been identified as such. This information is instead provided by the \emph{Recall} metric: a low \emph{Recall} means that many spambots are left undetected. \emph{Specificity} instead measures the ability to identify human users as such. Finally, \emph{F-Measure} and \emph{MCC} convey in one single value the overall quality of the prediction, combining the other metrics. Moreover, \emph{MCC} is considered the unbiased version of the \emph{F-Measure}, since it uses all the four elements of the confusion matrix. Being a correlation coefficient, $MCC\thickapprox1$ means that the prediction is very accurate, $MCC\thickapprox0$ means that the prediction is no better than random guessing, and $MCC\thickapprox-1$ means that the prediction is heavily in disagreement with the real class.

\subsection{Results and Discussion}
Table \ref{tb:result} shows the performance of our proposed neural model along with that of existing traditional techniques and algorithms reported on the cresci-2017 dataset. On test set \#1, our \emph{Recall} score outperforms the best, Cresci et al. \cite{cresci2016dna}, by 0.4\% (absolute). On test set \#2, our \emph{F-Measure} surpasses the best, Cresci et al. \cite{cresci2016dna} and Ahmed et al.\cite{ahmed2013generic}, by 0.3\% (absolute); the \emph{Accuracy} score is the same as the best, Cresci et al. \cite{cresci2016dna}. Most  of our other scores are comparable to those of existing work.

As shown in Table \ref{tb:result}, Davis et al. \cite{davis2016botornot}, Yang et al. \cite{yang2013empirical} and Miller et al. \cite{miller2014twitter} achieve rather unsatisfactory results for the test set \#1. The low values of \emph{F-Measure} and \emph{Mathews Correlation Coefficient} (MCC), respectively smaller or equal to 0.435 and 0.174, are mainly due to the low \emph{Recall}. In turn, this represents a tendency of predicting social bots as genuine accounts. As for the test set \#2, results in Table \ref{tb:result} show that Miller et al. \cite{miller2014twitter} achieved the worst performances among all those that we have benchmarked in this study. Low values of both Precision and Recall mean incomplete and unreliable bot detection. As reported in Table \ref{tb:result}, our approach proved effective in detecting Twitter bot, with an \emph{MCC} = 0.920 for test set \#1 and \emph{MCC} = 0.857 for test set \#2, comparing to the other four systems, i.e., Davis et al. \cite{davis2016botornot}, Yang et al. \cite{yang2013empirical}, Miller et al. \cite{miller2014twitter} and Ahmed et al.\cite{ahmed2013generic}.

Our model outperforms the current state-of-the-art algorithm by Cresci et al.~\cite{cresci2016dna} on several metrics such as accuracy and F-measure (on test set \#2) and recall (on test set \#1).  Although our model performs slightly below the algorithm in~\cite{cresci2016dna} on some other metrics, our model offers many significant advantages over~\cite{cresci2016dna}:
\\
\textbf{No handcrafted features required}:  Our model does not rely on any human-engineered features. The technique by Cresci et al., on the other hand, requires two large groups of (i.e., a set of six) user behaviour features and introduces a bio-inspired technique to model online user behaviors by so-called ``digital DNA'' sequences. The process of digital DNA fingerprinting has four main steps: (i) acquisition of behavioral data; (ii) extraction of DNA sequences; (iii) comparison of DNA sequences; (iv) evaluation. It is very time consuming and labour intensive to select and collect good handcrafted features. \\
\textbf{No prior knowledge required}: Our model does not require prior knowledge or assumptions about users’ profiles, friendship networks, or historical behavior on the target accounts. We only rely on the textual contents of users' tweets. The technique by Cresci et al., on the other hand, requires both tweet type feature and tweet content feature, and thus feature engineering.  It is expensive to collect, store and pre-process a large amount of data based on features.  Our model can avoid these costs.  Without feature engineering, our model can be implemented and deployed much faster and earlier than the other algorithms.

\begin{table*}[t!]
\caption{Performance comparison among various spambot detection techniques and algorithms reported on the cresci-2017 dataset.}
\begin{center}
  \begin{tabular}{c c c c c c c c}
  \hline      
  & & \multicolumn{6}{c}{\textbf{detection results}} \\
  \cline{3-8}      
\textbf{technique}  &   \textbf{type} &  Precision  & Recall  & Specificity & Accuracy & F-Measure & MCC \\
\hline      
test set \#1 \\
Human annotators        & manual & 0.267         & 0.080 & 0.921          & 0.698 & 0.123 & 0.001 \\
BotOrNot?\cite{davis2016botornot}  & supervised    & 0.471 & 0.208 & 0.918 & 0.734 & 0.288 & 0.174 \\
C. Yang \emph{et al.}\cite{yang2013empirical}  & supervised    & 0.563 & 0.170 & 0.860 & 0.506 & 0.261 & 0.043 \\
\hline      
ours                    & supervised   & 0.940	& \textbf{0.976} & 0.935 & 0.961 & 0.963 & 0.920 \\
\hline      
Miller \emph{et al.}\cite{miller2014twitter} & unsupervised  & 0.555 & 0.358 & 0.698 & 0.526 & 0.435 & 0.059 \\
Ahmed \emph{et al.}\cite{ahmed2013generic}  & unsupervised  & 0.945 & 0.944 & 0.945 & 0.943 & 0.944 & 0.886 \\
Cresci \emph{et al.}\cite{cresci2016dna} & unsupervised  & \textbf{0.982} & 0.972 & \textbf{0.981} & \textbf{0.976} & \textbf{0.977} & \textbf{0.952} \\
\hline      
\hline      
test set \#2 \\
Human annotators        & manual        & 0.647 & 0.509 & 0.921 & 0.829 & 0.570 & 0.470   \\
BotOrNot?\cite{davis2016botornot} & supervised    & 0.635 & \textbf{0.950} & 0.981 & 0.922 & 0.761 & 0.738   \\
C. Yang \emph{et al.}\cite{yang2013empirical} & supervised    & 0.727 & 0.409 & 0.848 & 0.629 & 0.524 & 0.287   \\
\hline      
ours                    & supervised & 0.933 & 0.919 & 0.938 & \textbf{0.929} & \textbf{0.926} & 0.857 \\
\hline      
Miller \emph{et al.}\cite{miller2014twitter}   & unsupervised  & 0.467 & 0.306 & 0.654 & 0.481 & 0.370 & -0.043  \\
Ahmed \emph{et al.}\cite{ahmed2013generic}  & unsupervised  & 0.913 & 0.935 & 0.912 & 0.923 & 0.923 & 0.847   \\
Cresci \emph{et al.}\cite{cresci2016dna} & unsupervised  & \textbf{1.000} & 0.858 & \textbf{1.000} & \textbf{0.929} & 0.923 & \textbf{0.867}   \\
\hline      
\end{tabular}
\label{tb:result}
\end{center}
\end{table*}

\begin{figure*}[t!]
    \centering
    \begin{subfigure}[t]{0.5\textwidth}
        \centering
        \includegraphics[height=5cm]{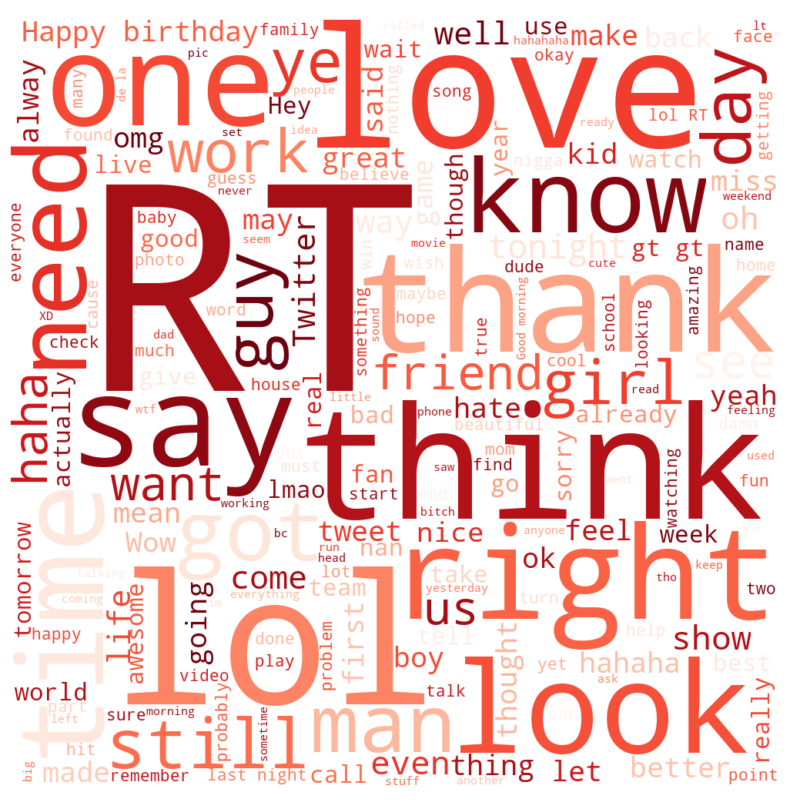}
        \caption{human account}
    \end{subfigure}%
    ~
    \begin{subfigure}[t]{0.5\textwidth}
        \centering
        \includegraphics[height=5cm]{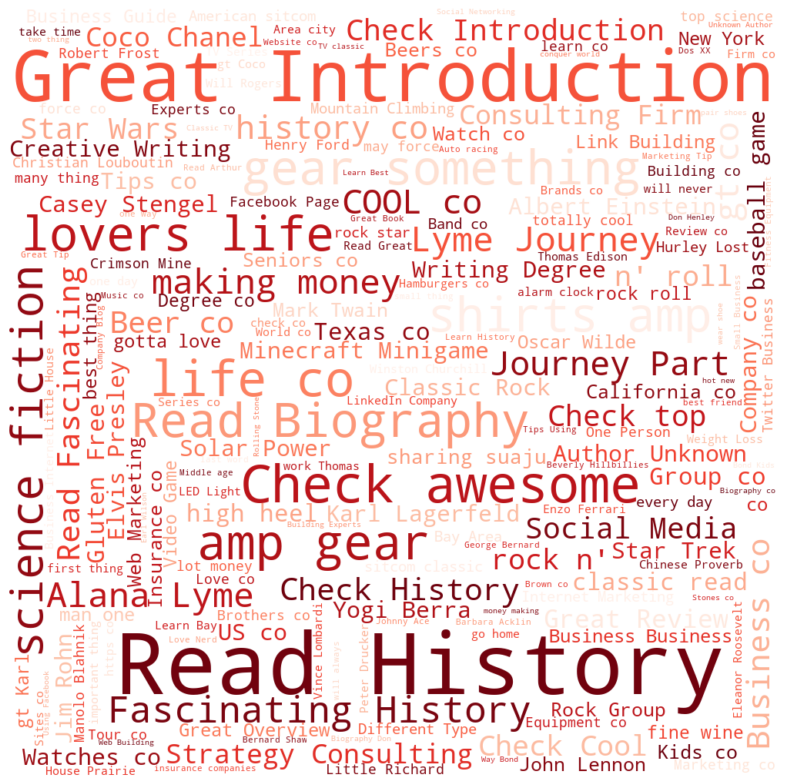}
        \caption{social-bot-3}
    \end{subfigure}
    \caption{Word-cloud comparison of human account dataset and social-bot-3 dataset.}
    \label{fig:WordCloud}
\end{figure*}

In order to gain more insight into the datasets, and thus the effectiveness of our proposed model, we generated a \emph{word-cloud} for comparison of the most frequent words in the two datasets, i.e., human accounts and social-bot-3, as shown in Fig.~\ref{fig:WordCloud}. Word-cloud is a visualisation method that displays how frequently words appear in a given body of text, by making the size of each word proportional to its frequency. It is worth noting that Amazon social bots on Twitter usually prefer to use exaggerated words such as \emph{Check awesome}, \emph{Read Fascinating} and \emph{Creative Writing}, to attract people's attention in order to advertise their products or services, or are talking about trends in a particular domain. Furthermore, a manual analysis of 100 randomly selected tweets in social-bot-3 showed that a majority of their tweets contains links to external web pages. This is in contrast to the general human accounts (in the random sample), which describe the accounts' owners using words such as \emph{love, happy birthday, haha, lol, thank}, and \emph{friend}, and most of whom seldom tweet links to external web pages.

Overall, the promising preliminary experimental results are yielded by the effective and efficient modeling ability of a deep bidirectional recurrent neural network architecture with the word embedding technique, as well as the availability of a large set of public annotated training data~\cite{mibdatasets} that we used in this paper. Our proposed model outperforms the state-of-the-art algorithm~\cite{cresci2016dna} in several metrics and has the significant advantage of not using any feature engineering or prior knowledge.  This will save time and money to select, collect, store and pre-process data.  This advantage also enables faster and easier implementation and deployment of a bot detection scheme in real life.

\section{CONCLUSION}

This paper presents a RNN model, specifically BiLSTM, with word embeddings to distinguish Twitter bots from human accounts.  Our model requires no prior knowledge or assumption about users' profiles, friendship networks, or historical behavior on the target account. To the best of our knowledge, our work is the first that develops a RNN model with word embeddings to detect bots that relies only on tweets and does not require heavy feature engineering. The preliminary simulation results are very encouraging. Experiments on the public dataset cresci-2017 show that our model can achieve similar performance compared with existing work, without handcrafted feature engineering, which is labor intensive and time consuming.  This advantage allows for faster and easier implementation and deployment of the bot detection scheme.
In addition, our proposed bidirectional recurrent neural architecture can be relatively easily adapted to a new problem, for example, using BiLSTM with word embeddings to detect phishing email, webpages or SMS.


\vspace{12pt}

\bibliographystyle{./IEEEtran}
\bibliography{./IEEEabrv,./IEEEexample}
\end{document}